%% file: paper.tex
\documentclass[conference]{IEEEtran}

\newcommand{\scomment}[1]{}   

\usepackage{amsmath}
\usepackage{amssymb}
\usepackage{subfig}
\usepackage{paralist}
\usepackage{wrapfig}
\usepackage{multirow}
\usepackage{graphicx}
\usepackage{mdwlist}
\usepackage{url}

\title{Blindspot: Indistinguishable Anonymous Communications}

\author{\IEEEauthorblockN{Joseph Gardiner and Shishir Nagaraja}
\IEEEauthorblockA{School of Computer Science\\
University of Birmingham\\
Birmingham, UK, B15 2TT\\
Email: j.gardiner, s.nagaraja@cs.bham.ac.uk}}

\begin{document}
\maketitle

\input{introduction}

\input{problem}

\input{designanalysis}
\input{results}

\input{analysis}

\input{background}

\input{conclusion}

\bibliography{paper}

\bibliographystyle{abbrv}

\end{document}

%% file: introduction.tex
\begin{abstract}

Communication anonymity is a key requirement for individuals under
targeted surveillance. Practical anonymous communications also require
indistinguishability --- an adversary should be unable to distinguish
between anonymised and non-anonymised traffic for a given user. We
propose Blindspot, a design for high-latency anonymous communications
that offers indistinguishability and unobservability under a
(qualified) global active adversary. Blindspot creates anonymous
routes between sender-receiver pairs by subliminally encoding messages
within the pre-existing communication behaviour of users within a
social network. Specifically, the organic image sharing behaviour of
users. Thus channel bandwidth depends on the intensity of image
sharing behaviour of users along a route. A major challenge we
successfully overcome is that routing must be accomplished in the face
of significant restrictions --- {\em channel bandwidth is
  stochastic}. We show that conventional social network routing
strategies do not work. To solve this problem, we propose a novel
routing algorithm. We evaluate Blindspot using a real-world
dataset. We find that it delivers reasonable results for applications
requiring low-volume unobservable communication.
\end{abstract}

\section{Introduction}

Anonymous communication networks such as Tor and JAP~\cite{tor,JAP}
have, in a short period of time, revolutionised communication security
landscape on the Internet. These systems and others~\cite{DDM03,drac}
provide anonymous communications which can defend against traffic
analysis attacks. Often, an adversary will try to build a dossier of
the most central nodes, and attacking those on the top of the
list. Traffic analysis of inter-node communication offers basic tools
to collect necessary intelligence in order to plan an attack --- i.e
an understanding of their communication traffic behaviour --- for
instance, who they talk to, how often, and the frequency and timing of
communication (all of this is contained in metadata, which is
collected, and stored, by GCHQ as part of Tempora).

This information is inferred from the traffic data at appropriate
vantage points on the Internet. A map of internet backbone links shows
obvious positions where large percentages of internet traffic pass
through. For example, a majority of the traffic between the USA and
Europe passes through Cornwall, where GCHQ also happens to have an
listening station at Bude. According to the Snowden revelations, this
station captures all traffic passing over ten of these undersea cables
as part of the Tempora program. This justifies the need to consider
global active adversaries in our threat model.
 
In anonymous communication networks, anonymity is provided under the
assumption that users solely communicate via anonymous channels. This
is unrealistic as widespread adoption of anonymous communications is
hindered by usability problems (performance overheads) as well as lack
of awareness. Therefore, users of anonymous communication networks
also have to communicate over conventional channels. Thus the
statistical characteristics of the anonymous channel of a user must be
{\em equivalent} to (or {\em indistinguishable } from) non-anonymous
communication channels. {\em Indistinguishability} is the property
that anonymous and non-anonymous communications are not differentiable
by an adversary. Indeed, {\em without indistinguishability,
  anonymous communication networks have a basic usability
  problem}. 

Current approaches to this challenge take two directions. One line of
work is on censorship resistance mechanisms. These
works~\cite{Burnett:2010:CAC,Invernizzi:2013:MBS,huang:08:stegviop,rowland:97:ip,ahsan:02:tcpip}
all introduce extra communication endpoints (such as cloud file storage
services) into traffic, meaning that there is some form of
distinguishable behaviour to monitor. Approaches that make use of
protocol-mimicking (attempting to hide one type of traffic by making
it appear as
another)~\cite{moghaddam:2012:skypemorph,wang:2012:censorspoofer,weinberg:2012:stegotorus}
fall victim to incomplete emulation of the protocols that they are
attempting to mimic~\cite{houmansadr:2013:parrot}. 

A second direction is the design of anonymous communication networks.
The unlinkability property~\cite{PH00} alone is inadequate for a
number of reasons. First, as Danezis et al. pointed out, it does not
hide the volume of communications and hence leaks enough information
to a global adversary who can compile an ordered list of
targets~\cite{danezis:2006:weis}. Second, it does not defend against
traffic confirmation attacks~\cite{RSG98, statistical-disclosure}
where an adversary injects traffic load patterns to determine
communicating end points. To combat these problems full
unobservability --- passive attacker cannot distinguish whether or not
a user is communicating --- is needed. Systems such as
Drac~\cite{drac} provide unobservable anonymous
communications. However, the attacker can still distinguish between
the following communicating states of a user: whether or not
'unobservable' anonymous channels are in use, so the protection
accorded is still short of what users need.

\subsection{Motivation} Without indistinguishability, the use of
anonymous communications alone could incriminate a user simply by
identifying their use. The use of anonymous communication channels can
cause escalations such as a targeted intrusion attack on the user or
worse, trigger a physical attack on the individual. Anonymity without
company~\cite{DM06}, can make identification of a user of anonymous
communications much easier. For example, in 2013 a student at Harvard
University sent an bomb threat through email in order to get his
imminent economics exam cancelled~\cite{torbombthreat}. The student
used Tor to send an email via the Guerrilla~\cite{geurrilla}
disposable email service, assuming he would be unidentifiable. The
investigators simply compiled a list of Tor users filtered by the
timing of the threat to identify the student.

\subsection{Research problem} Credible defences against targeted surveillance
attacks require both indistinguishability and unobservability. This
places significant restrictions on the statistical characteristics of
user traffic --- the traffic shape should be independent of the
presence of anonymous communications.

To address these requirements, we propose {\em Blindspot}, an
anonymous communication network that leverages conventional
communication channels on social networks. Specifically, the
image-exchange behaviour of users. In Blindspot, nodes communicate by
broadcasting messages to their neighbours. Messages are
steganographically embedded within an image the user uploads. When
uploaded by the communicating (sender) node, the message carrying
image is available to all its neighbours. Each participating node
checks for incoming messages by monitoring images uploaded by its
neighbours. It provides {\em indistinguishability} by ensuring that
the image upload behaviour of participating users remains unchanged
and provides probabilistic {\em unobservability} through the use of
steganography.

To the best of our knowledge, Blindspot is the first system to provide both
properties. This may make it a useful building block of some forms of anonymous
communications, such as announcing a meeting of a social club (corporate gathering, undercover organisation, dissident organisation, or a protest group).

Blindspot achieves these properties through the use of social trust
relationships within the routing process. Unlike previous proposals based on
steganographic techniques~\cite{Burnett:2010:CAC,Invernizzi:2013:MBS,huang:08:stegviop,rowland:97:ip,ahsan:02:tcpip}, which introduce additional
communication endpoints, Blindspot routes through the pre-existing social network, thus exploiting trust
relationships to secure routing. We leverage this in four ways. First, all
routing is through pre-existing social fabric interconnecting nodes, therefore no new communication endpoints are introduced. Second, it leverages the diffusion
properties of the social network to efficiently and anonymously route to the
destination. Third, Blindspot does not alter statistical characteristics of
conventional user traffic. This is achieved by piggy-backing on existing image
sharing behaviour, no extra images are added nor are extra uploads
scheduled. Channel traffic characteristics, as observable to the adversary, are
not altered. So routing depends wholly on the innate image sharing behaviour of
the users. This is a non-trivial challenge. Blindspot operates on a network
topology where channel bandwidth is severely constrained. And, the latency is a
function of stochastic user behaviour along the entire route. Fourth, social
networks can be used to detect Sybil attacks~\cite{sybillimit,sybilguard,sybilinfer,xvine} by detecting a small cut that separates honest
and sybil groups. Blindspot benefits from Sybil resistance properties of social
network topologies. Specifically, this restricts the power of adversaries that
seek to inject large numbers of misbehaving participants into the network.

We evaluate Blindspot empirically using real-world social network topologies and
real user behaviour. From our evaluation, we find that Blindspot is able to
route $90\%$ of messages within a day across
diverse routes. We also validate our work on a synthetic social network
topology.

The structure of this paper is as follows. First, in Section~\ref{sec:problem} we will give a more formal definition of the problem. This will be followed by an overview of the system design (Section \ref{sec:overview}), and then a full evaluation of the systems performance in Section~\ref{sec:results} and discussion of the security provided by this system in Section~\ref{sec:analysis}. Finally, we will give an overview of existing systems in Section~\ref{sec:background}.

%% file: problem.tex
\section{Problem Description} \label{sec:problem}

Blindspot's goal is to offer high-latency anonymous communications with the following properties:  \begin{inparaenum}[(a)]
\item {\em Indistinguishability} between (high latency) anonymised and non-anonymised communications of a given user. 
\item {\em Unlinkable} communication between sender and receiver pairs.
\item {\em Low-delay any-to-any routing} in a network with {\bf stochastic node
  behaviour}.
\end{inparaenum}

In this situation low-delay is used in relative terms. The system will not be able to achieve latencies that match that of systems such as Tor, rather message delivery can be measured in hours or even days. The low-delay property represents the routing algorithm's design that aims to reduce this delay as much as possible.

\subsection{Low-delay routing in a capacity constrained network}
Consider a social network represented by a graph $G=(V,E)$, where nodes are
people and edges are trust relationships. Each node $v_i \in V$ in the graph
is associated with a probability distribution over time elapsed between images
posted by the corresponding user. The time between image postings is the minimum amount of time that a message can spend sitting on a node. This is called the {\em delay
distribution} of $v_i$. 

A route of length $k$ on the network is a sequences of vertices $v_1,v_2,\dots,
v_k$. A message sent along this route will be influenced by the delay
distribution of {\em all the nodes} along the route. We assume that a node
downloads all messages broadcast by its neighbours whenever it uploads. This
assumption is based on the current operational behaviour of many OSNs (which 
predicatively cache image posts from neighbours for performance reasons).

The problem of routing in a network with weighted nodes, where each weight is an
independent random variable, is an unsolved problem. Note, that graph weights
are not generated from a probability distribution function. Instead, each node
is associated with {\em a separate PDF} which is different (and independent) of
all other nodes in the graph.

\subsection{Indistinguishability} 
In a nutshell, indistinguishability is defined as the inability to distinguish
between anonymous and non-anonymised traffic traces of a specific
user. Blindspot leverages photo exchange traffic of users on social networks as
the carrier channel. We formally define the Indistinguishability Game (inspired
by Cryptography literature), played between the attacker and a communicating
user: First, consider photo upload traffic $P^{X}_0$ generated from a
distribution over inter-upload delay times of the user $X$, when the user {\bf
  is not} communicating via Blindspot. Next, consider the photo upload traffic
$P^{X}_1$ when the user {\bf is} communicating via Blindspot. Now a random $i
\in {0,1}$ is generated by a fair coin flip, the defender is given one or more
click traffic sets from $P_i$ and he/she outputs $i'$. The attacker wins the
game if $i=i'$. A communication channel is `indistinguishable', if no attacker
can win the Indistinguishability Game with probability non-negligibly greater
than $0.5$.

\subsection{Threat model}

The Blindspot system attempts to achieve its goals under two types of adversaries. Firstly, the primary threat as considered by this system is the global active adversary. This refers to the OSN provider (for example, Facebook), who we assume has access to all images uploaded to the network and the full underlying social graph. We also assume that the OSN has the capability to modify images in any way that they see fit. The OSN may be colluding (not necessarily knowingly) with an external surveillance agency (for example the NSA or GCHQ). 

Secondly, we consider the partially active insider. We assume that the adversary can insert dishonest nodes, either through the introduction of Sybil nodes or by compromising existing users. These nodes can be used to monitor traffic on the network, by collecting any images that pass over them, and has access to the embedded data. The insider can perform tagging attacks on the messages to attempt to remove unlinkability, and can instruct insiders to perform DOS attacks through blackholing (refusing to route messages), or by flooding messages in order to disrupt routing. 
Blindspot offers partial defences against a subset of active insider attacks, but not all of them. A detailed discussion on the system's resistance and susceptibility to various attacks can be found in Section~\ref{sec:analysis}.

\subsection{Intended Use}
The system in its earliest form will only support the sending of short, text based messages (around the length of a tweet), including anything that can be encoded into this length of text as hexadecimal or base64 strings. The message data could undergo external compression in order to maximise capacity. For data that is too long to send, multiple messages could be sent provided the contents contain error coding (to mitigate undelivered portions of the data), or the message could consist of a pointer to the data at an external source (for example a link to a file on a cloud storage service).
 
While short messages may not seem useful, in many cases it is enough in order to communicate key points. For example, in the case of a protest group organising a protest, the organiser simply needs to send a time and location. A whistleblower only needs to send the details of either a dropzone for documents or a location to arrange a meet. 
\subsubsection{Closed Networks}
The system supports the ability for users of the system to create smaller, closed networks in order to provide communication solely within a social group. For certain users groups, being able to separate from general traffic and tweak system parameters to suit the situation could provide extra benefits. The use of many smaller, more closed networks has the benefit that there is less risk of congestion due to smaller user numbers, and better trust relationships in order to resist insider attacks. This is discussed more in section~\ref{sec:overview}.

%% file: designanalysis.tex
\section{System design}
\label{sec:overview}

The major component of the design of Blindspot is the new probabilistic routing algorithm that is able to successfully route within the confines of the image sharing behaviour on an online social network. To the best of our knowledge, routing across nodes who are communicating at natural rates is an unresolved problem. 
\subsection{Basics}

\begin{table*}
\centering
\small
\begin{tabular}{@{\extracolsep{1em}}ll@{}}\hline
$N_{r}$   & Unique nonce to identify message\\
$(K^{M}_r, P^{M}_r) = (y = g^x ,x)$ for $x \in_U \mathbb{Z} _q$. & Message public-private encryption key-pair of receiver\\
$(K^{N}_r, P^{N}_r)$ & Neighbourhood public-private encryption key-pair of receiver\\
$m$      & Communication message\\
$K{m}$  & Elgamal encryption of m under key K\\
$K[m']$ & Re-encryption of encrypted message $m'$\\
$[K^N_r{N_r, K^M_r{M}}]$ & Message structure\\\hline
\end{tabular}
\caption{Terminology}
\label{tab:terminology}
\end{table*}
\normalsize

\subsubsection{Message-secrecy key pair} Each participating node generates a public
 encryption key and a private decryption key according to a public-key
 cryptosystem. A pair of communicating nodes (sender and receiver) exchange
 public encryption keys out-of-band. For example, receivers could publish public keys on an external site or distribute in person in a similar fashion to how PGP keys are distributed today.

\subsubsection{Neighbourhood key pair} Each node also generates a second encryption key-pair, called the neighbourhood key-pair. The private encryption key is shared with all of the node's neighbours. This allows them to deliver intended messages directly to the receiver, and stop forwarding the message any further.

\subsection{Routing design}
At the core of Blindspot is fully decentralised routing algorithm, based upon a pull--broadcast model. Messages are routed between a sender and receiver without revealing receiver identity to
any intermediate node on the route, except the last `exit' node.  

Each node on the social network maintains an input message queue (new incoming
messages) and an output message queue (outgoing messages).

\subsubsection{Node Knowledge}
We make the assumption that every node on the network has a full knowledge of only those nodes for which it is directly connected, i.e. their ``friends''. This knowledge includes the image upload behaviour of those nodes, and their friend lists. This is the typical case for many social networks. For example, on Facebook (by default) a user can view all activity performed by a friend (posts are by default visible to friends and friends of friends), and can, usually, see the friends list of that friend. The exception to this is if the friend has set extra privacy setting on their account, but this is the small minority of users.

A node has no knowledge of the global network topology, only the 2-hop  neighbourhood of which it is the centre node (the node, its friends and its friends' friends). Outside of this 2-hop neighbourhood, the node has zero knowledge of the graph topology or the location and behaviour of other nodes. Each node only has the ability to communicate with (share images with and pull images from) its direct neighbours.

\subsubsection{Message construction}
 
Each message is of the form $K^N_r{N_r, K^M_r{M}}$. To create a new message,
the sender encrypts the message with the receiver's message-secrecy public key. This ensure message privacy as only the receiver can recover the actual contents of a message. Next, it generates a unique nonce $N_r$. It then encrypts the the nonce to provide a function for nodes in the receiver's neighbourhood to discard duplicates, and for the receiver to recognise duplicate messages. 

The receiver's address (this will vary according to the social network, but will typically be the username) is encrypted with the receiver's neighbourhood key under the Elgamal-based scheme described in section~\ref{sec:tagging} using the receiver's neighbourhood public key.

The user agent running on the node now places a newly constructed message and encrypted destination address into the node's output message queue.

Each node on the path therefore only sees the encrypted destination and message. Each node, due to the use of re-encryption, will see a different ciphertext for each component, providing unlinkability and resistance to tagging attacks. An important point is that different paths that a message takes result in a different series of ciphertexts. The nodes who possess the receiver's neighbourhood key will be able to identify the destination of messages for them, but will be unable to view the message contents.

\subsubsection{Output queue processing}
To process messages on the output queue, the user-agent waits for the user to
upload an image. Note: the {\bf same image is not used} for each upload. The
image selection and uploading is carried out by a human, not the user-agent. The
state of the message queue has no impact on the user's decision to post an
image.

Once the user selects an image, the user-agent intercepts it and
steganographically embeds messages into the image. The precise method of doing
this in a reliable manner without interference from OSN compression algorithms
is given in Section~\ref{sec:stego} - ``Channel design for unobservability''.

\subsubsection{Input queue processing and routing strategy}
Similarly, when the user reads an image post, as part of their normal browsing behaviour, the image is downloaded to the
user's computing device. The user-agent then extracts embedded message within
the image and enters them into the input queue. Blindspot then selects a subset
of messages from the input queue which are moved to the output queue according
to a routing strategy.

Traditionally, the path that is taken between two endpoints is usually taken to
be the shortest between them in the number of hops or distance, which is the assumption that many traditional routing and path selection algorithms work to.  In this situation, however,
the shortest path in terms of hops may not necessarily be usable if not all of
the nodes on the path are active (or frequent) uploaders. The state of the network in terms of
which nodes are ``active'' also changes often, meaning traditional shortest path
algorithms are insufficient (again, these usually work on the assumption that nodes are static in behaviour).

We therefore use the inter-upload distributions for each node to aid in the routing decisions, such that the chosen paths represent the lowest waiting time (time between message upload and subsequent upload by neighbour) resulting in a relatively low delay path.

The upload delay distribution of each node is governed by a random variable. Subsequently, the distribution over delivery times, or the message-delay distribution is dependent on this. For any given route, the message-delay distribution is the sum of independent random variables (of nodes) on that route. For a two hop route (s, t, r) where s (sender) is communicating with r (receiver) via t (intermediary), the message delay distribution is given by D = S + T, where S and T are random variables whose PDFs are the inter-upload delay distributions of nodes $s$ and $t$ respectively.

The sum of two random variables is equivalent to the convolution of the respective PDFs, given by the convolution formula:
\begin{equation}\label{eq:convolve}
(f * g )(t) {=}\ \int_{-\infty}^\infty f(\tau)\, g(t - \tau)\, d\tau
\end{equation}
By applying this formula to every pair of node on the path, the result will be the distribution of the possible total delay of the message. In our system, however, it is not possible to compute this full convolution as part of the routing algorithm as the convolution output (which increases in size by the length of the additional distribution at each convolution) will reveal the number of hops the message has taken if it is passed along with the message (it will also use up a large chunk of capacity). An example convolution output can be found in Figure~\ref{fig:cdfscore}(b).

To solve this problem, we instead use the local knowledge of nodes in order to compute the convolution a small section of each path, without passing any values along with the message.

As we have said, each node knows the image upload distributions of its neighbours (plus itself). This is enough knowledge to compute a two hop section the path (the previous node, the current node, and the next node). For a node with degree $d$, this equates to $d*(d-1)$ possible paths (from all neighbours to all neighbours excluding the source). 

\paragraph{Path selection} A node cannot pull every message that is sees due to the capacity limitations. Therefore nodes need to make some decision on which messages to pull and forward. We use a probabilistic approach for deciding this. 

During an upload session (the time period in which the user is making uploads, where 1 or more images will be uploaded) the node will fetch the messages contained in the last 5 uploads made by each neighbour (limited by a TTL) that they have not already processed. Note that uploads can be fetched over time during normal browsing behaviour and the messages contained cached. This means that messages can be processed in the background over time, including the more costly operations such as decryption and re-encryption. It also means that there is not a sudden spike in requests for neighbours images, which could be an indication of the systems use. The messages from all of these uploads is placed in a queue, which is then sorted according to a score function, given by eq.~\ref{eq:mscore}, where $similarity_{prev}$ is the percentage of neighbours shared by the current node and the neighbour the messages was received from (the lower this value the better as it allows the message to escape communities), and $min(cdf)_{src}$ is the minimum convolution score for all possible exit routes (all neighbours excluding the node from which the message was received) (see section~\eqref{sec:convolution} below, ``Convolution Comparison'') normalised between 0 and 100, which indicates the future upload prospects of the message, a lower score indicating the greater likelihood of delivery. 

The score of message $i$ is given by:
\begin{equation}\label{eq:mscore}
score_{m_i} =  similarity_{prev} + min(cdf)_{src}
\end{equation}

\paragraph{Message expiration} The messages are sorted in ascending order according to this score (eq.~\ref{eq:mscore}, lower scores are desirable). The node will then go down this list, and perform a weighted coin toss for each message to decide if it should be pulled --- moved from the input queue to the output queue. Each message is selected with a probability $1/deg_{src}$, or discarded with probability $1- 1/deg_{src}$, meaning that the lower the degree of the proceeding node, the more likely the message is to be forwarded. Over time this provides an natural expiration effect, message sprays will thin as time passes.

\subsubsection{Convolution Comparison}\label{sec:convolution}
We take the convolutions of the previous node, the current node, and each possible next node (resulting in the $deg_{current node}-1$ distributions). Each of these are then converted into a cumulative density function (CDF), and a score is computed for each. The score measures the steepness of the CDF at all points along it, such that a lower score represents a CDF that climbs (and reaches 1) early. The score is as follows:
\begin{equation}\label{eq:cdfscore}
\sum_{i=1}^{length(cdf)} (1-P_{i})\cdot X_{i}
\end{equation}
where $P_{i}$ is the CDF value at position $i$, $X_{i}$ is the X axis value at position $i$ and $length(cdf)$ is the number of values along the X axis. We take $1-P_{i}$ to reward CDFs that climb early, as a CDF that reaches 1 sooner will have more of its values multiplied by zero, resulting in a lower score. 

\begin{figure*}[!ht]
\centering
  \subfloat[Top score 398.1211, middle 94056.63 and bottom 226694.9]
                { \includegraphics[width=0.5\textwidth]{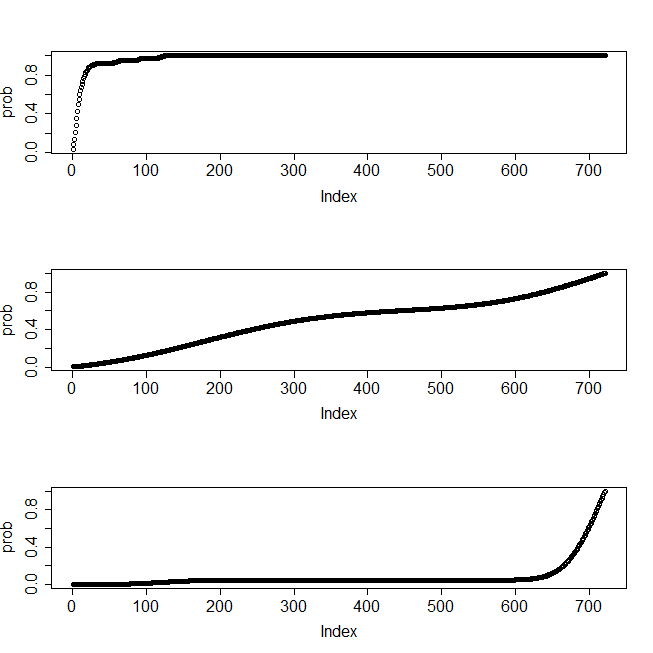}}
 \subfloat[CDF of convolution of PDFs in (a). Score is 618401.3]
         { \includegraphics[width=0.5\textwidth]{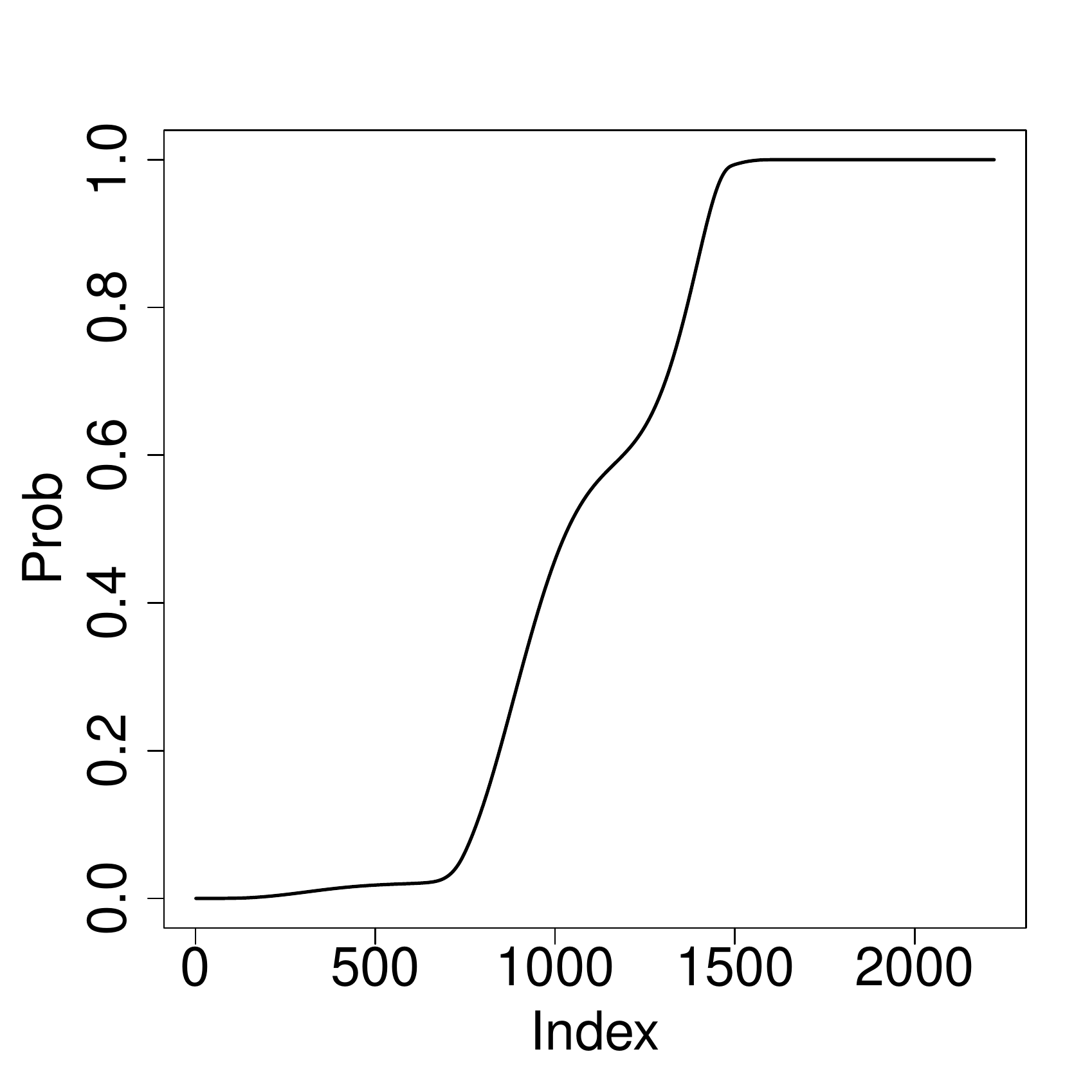}} 
 \caption{Score output for the CDF of three inter-upload PDFs and convolution output}
\label{fig:cdfscore}
\end{figure*}

A CDF with a lower score represents a PDF that is centred over a low delay, meaning that the path segment represented by it will likely have a low delay associated with it. Three example CDFs and their associated scores can be seen in figure~\ref{fig:cdfscore}.

The score will be larger for longer CDFs. If two CDFs of different lengths need to be compared, the shorter CDF must be extended with the addition of ones to the CDF values (or zeros to the underlying PDF values) to make them of equal length.

\subsubsection{Sybil Protection} Blindspot does not propose a new Sybil protection mechanism, but enjoys the benefits of existing techniques proposed in this space~\cite{sybillimit,sybilinfer,sybilguard,xvine}.

\subsubsection{Tagging attack resistance and unlinkability}\label{sec:tagging}
 To ensure that the input and output messages from each node are bitwise unlinkable and hence resistant to tagging attacks, we make use of universal re-encryption. As proposed by Golle et al~\cite{Golle02universalre-encryption}, universal re-encryption is a scheme that makes use of the ElGamal asymmetric key encryption algorithm to re-encrypt a message without the knowledge of recipient public keys. This is summarised as follows:\\
 \emph{Encryption of $m$ using key $K$:} $K{m}: C
= [(\alpha_0, \beta_0);(\alpha_1, \beta_1)] = [(my^{k_0}, g^{k_0});(y^{k_1},
  g^{k_1})]$ where $k_0, k_1 \in_U \mathbb{Z} _q$ and $q$ is a global system
parameter, which is 1024 bits in size.\\ \emph{Decryption of a given
  ciphertext:} $K^{-1}{(\alpha_0,\beta_0); (\alpha_1,\beta_1)}: m_0 =
\alpha_0/\beta_0^x$ $m_1 = \alpha_1/\beta_1^x$. If $m_1 = 1$ then the plaintext
is $m = m_0$.\\ \emph{Re-encryption:} $C'= [(\alpha_0', \beta_0');(\alpha_1',
  \beta_1)'] =
[(\alpha_0\alpha_1^{k_0'},\beta_0\beta_1^{k_0'});(\alpha_1^{k_1'},\beta_1^{k_1'})]$,
where $k_0', k_1' \in_U \mathbb{Z} _q$

At each point along the path a message takes, the forwarding node performs re-encryption on both ciphertexts. Any node on the path can perform this step with no knowledge of the receivers public key or the path that the message has taken/will take. Note that the receiver address and message are re-encrypted individually. 

The use of re-encryption provides sender-receiver unlinkability along all paths that the message takes, with no extra key setup apart from the initial receiver's public key distribution. This is more appropriate in this setting than the traditional onion routing approach, whereas the public keys for nodes on the path need to be collected by the source. Even if the path was known to the source, the key set-up stage would take far too long if key negotiation occurs within the system.

\subsubsection{Channel design for unobservability}\label{sec:stego}
There are two main requirements for the steganography algorithm used. First, and most importantly, the algorithm should be statistically undetectable~\cite{cachin98}, which means that no statistical test can distinguish between the set of cover objects and the set of stego objects. Secondly, in order to maximise the bandwidth within the system the algorithm also needs to ensure as much capacity is available as possible while maintaining data integrity in the face of compression. 

The system only requires that a group of users (such as one social group) agree on one steganographic scheme to use, which provides the trade-off between capacity and security to meet their needs. For particularly sensitive information, using an scheme which provides good security guarantees but a lower capacity will be preferable. For more general users, an higher capacity but lower security scheme could be used. 

An suitable example is the  YASS (Yet Another Steganography Scheme) algorithm for JPEG
steganography~\cite{Solanki07yass:yet}. YASS is an example of minimal distortion
embedding~\cite{fridrich05}~\cite{camenisch07}. Schemes that employ minimal
distortion embedding focus on increasing the embedding efficiency by decreasing
the embedding distortion. The YASS algorithm requires two parameters. $Q$
represents the quality factor and dictates the amount of compression performed
by YASS, and $q$ denotes the amount of redundancy provided by the algorithm. The
higher the value of $q$ the less data can be inserted into an image, but the more resilient the steganography to error. By default to provide unobservability YASS hides data at a rate of 0.05 bits per non-zero DCT coefficient.
Nagaraja et al~\cite{nagaraja2011stegobot} demonstrated the ability of YASS to withstand uploading to and downloading from Facebook. They found that, assuming a 720px image and a q value of 2, 40280 bits of information could be hidden (when tested on 116 different images). At the other extreme, with a $q$ value of 20, 4028 bits can be embedded. A sensible choice of $q=4$ results in 20140 bits of embedding capacity. The YASS algorithm also achieves a low bit error rate (BER, the ration of error message bits to the total number of bits hidden). Using a quality factor of $Q=65$ and redundancy parameter of $q=4$ results in an average BER rate of 0.1320. The can be reduced to 0.0003 with parameters $Q=65$ and $q=20$. These figures can be improved as some images are inherently ``bad'' (result in a high BER), by not making use of these images performance can be improved.
 
Fridrich et al~\cite{Fridrich:2007:SUJ} showed that both the MM3~\cite{camenisch07} and PQt~\cite{fridrich05} algorithms are also statistically undetectable at an embedding rate of 0.05 bits per non--zero DCT coefficient.

Facebook currently supports images of a maximum size of 2048px (2048*2048)
~\cite{facebookimage}.  Flickr also supports images of this size. This increase
in image size is of a great benefit to image steganography as the larger an
image is the more data can be hidden within it, we have estimated that this
increase in size can provide up to 8 times more capacity in an uploaded image
than as tested by Nagaraja et al. This gives a effective capacity of
8-10kb (assuming YASS with $q=4$. 

\paragraph{Steganographic Keys}
Each smaller network within the Blind system will be responsible for deciding upon which steganography algorithm is used. All members of the group can be expected to use the same algorithm. Each user will also have to generate an key which the will use when embedding the data. This will need to be given to neighbours, and can be shared along with the neighbourhood keys as described previously. Allowing each user to choose their own key will provide an extra level of security and prevent an insider leaking the key to extract the embedded data across all images. If a key is leaked, it will only affect the images of a single user, and can be replaced.

\subsection{Global and local networks}
As we have mentioned, as the system is fully decentralised it supports the creation of individual instances of the system. Within each instance, users will have to set up keys and decide upon a steganography algorithm. The main benefit of allowing these smaller networks is that it strengthens the trust model within the system. If the system is closed to users within a single social group, it makes it far more difficult for an unknown user to participate (as in a Sybil attack) and fewer users reduces the likelihood of user compromise due to fewer targets. 

%% file: results.tex
\section{Results}\label{sec:results}

To test the effectiveness of the routing algorithm we have produced a simple simulation in the Java language. Two main experiments are performed over two different networks. First, we test the performance of the routing algorithm under increasing congestion. We then test the effectiveness of the algorithm to maintain performance in the face of an active attack, namely the removal of increasing percentages of random and then high degree nodes (an effective black-holing attack). 

The simulator takes as input a graph and the individual upload behaviour of each node. We then choose pairs of nodes to communicate, such that a source sends to a destination.  The simulator works in a scale of days, with monthly image upload counts for each node split evenly amongst 30 days, for a total of 64 months. Each pair of nodes exchange 1 message per month. To keep the simulator lightweight, we do not apply encryption. After initial experimentation, we set the message TTL to 15 days. As per our channel design, assuming uploaded images are 4096px in size, we allow for 150 messages per image upload. We choose an image size of 4096px as it is (unofficially) accepted by Facebook for image uploads, and most modern smartphones feature a camera capable of taking images of this quality (or greater).

\subsection{Test Data}
The primary dataset that we use is the Flickr dataset from Nagaraja et al~\cite{nagaraja2011stegobot}. This dataset contains the social graph and monthly image upload behaviour of 7200 users of the Flickr social network. We use this dataset as it is increasingly difficult to apply crawling to social networks such as Facebook that include strict privacy settings and Twitter which has restricted the sharing of tweet data. The difficulty with sites such as Facebook is that, increasingly, profiles are set to private for everyone other than the friends of that person. While this would not impact on the systems ability to function, it prevents automated crawling of the social network. To calculate the inter-upload time for each node, we divide each monthly value by 720, which gives the estimated inter-upload time in hours.

To test the performance of the algorithm on a differently structured network, we also use a network generated using the Barabasi-Albert (BA) model~\cite{barabasi:2002}, with a parameter of 5. The Barabasi-Albert model produces a scale free network. We generate a network of 7200 nodes and apply the Flickr upload counts to this in a 1:1 mapping. We apply the 1:1 mapping as the resulting correlation between degree and upload count is very similar to that of Flickr.
\subsection{Effects of Congestion}
\begin{figure*}[!ht]
\centering
  \subfloat[Effects of congestion]
                { \includegraphics[width=0.5\textwidth]{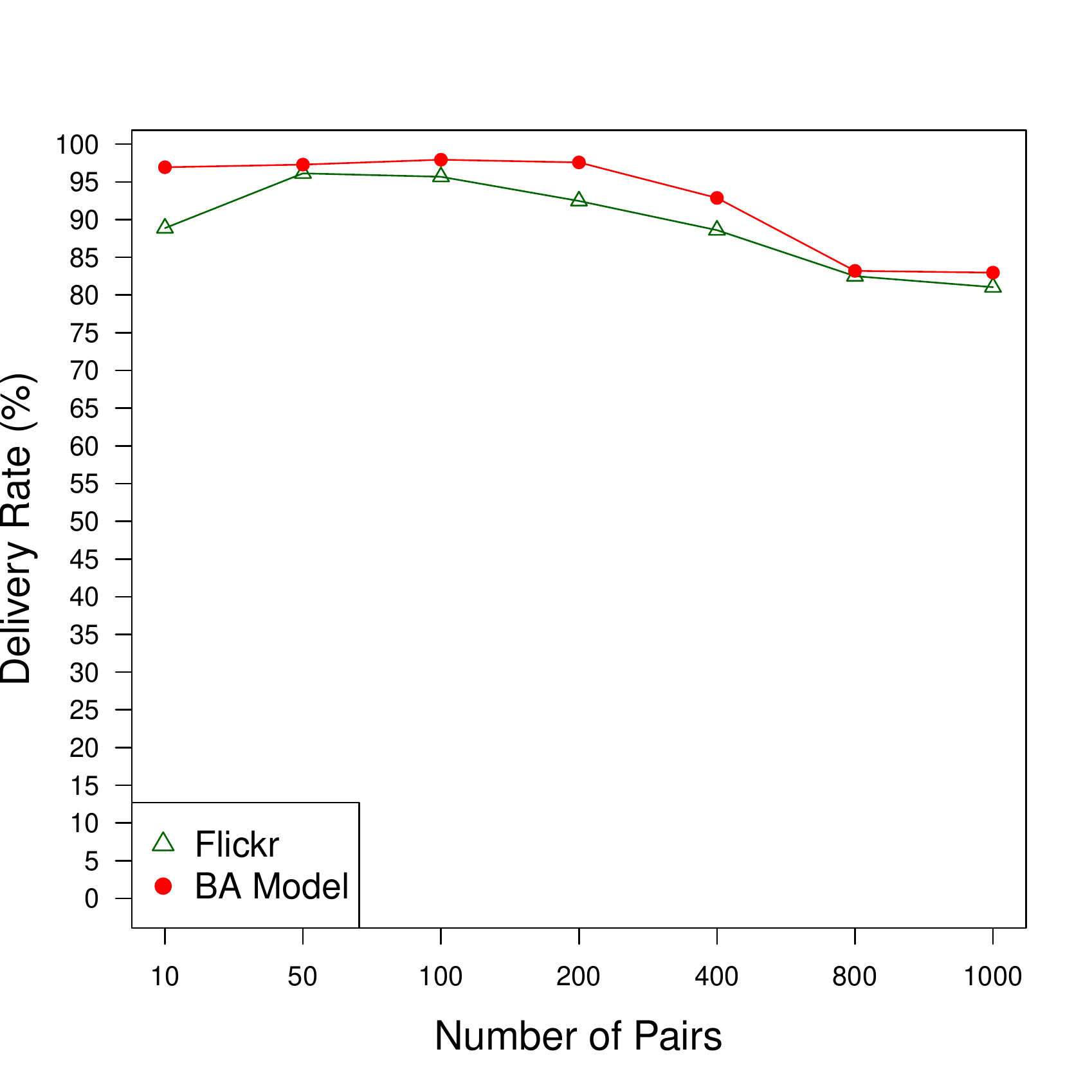}}
 \subfloat[Effect of black-holing both random and high degree nodes in the two test networks]
         { \includegraphics[width=0.5\textwidth]{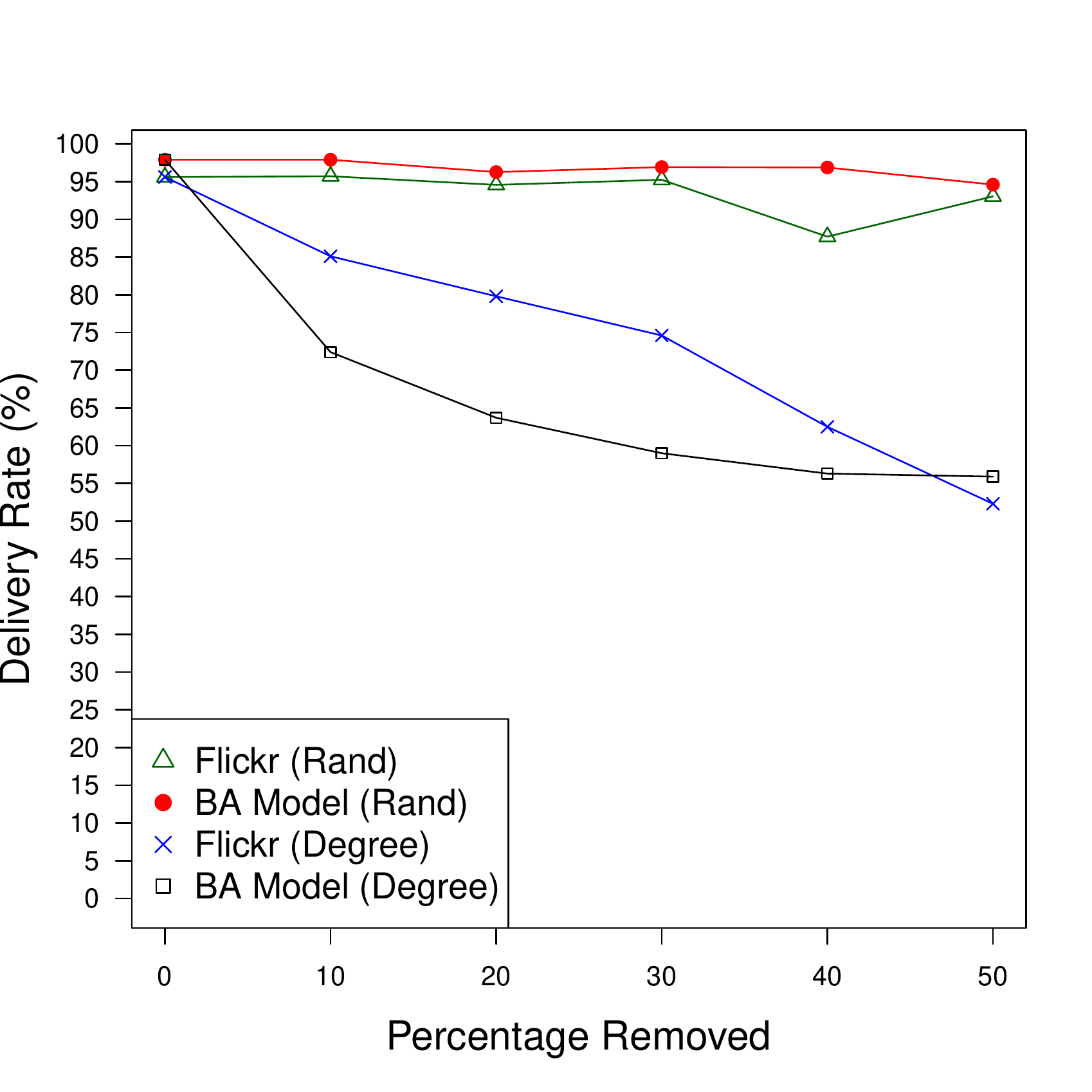}} 
 \caption{Performance under congestion and attack}
\label{fig:performancey}
\end{figure*}

In a capacity restricted network it is important to measure the effects of congestion. To measure the effects of congestion, we choose increasing number of pairs of nodes to communicates with 1 message per month per pair. The results of this experiment are presented in Figure~\ref{fig:performancey}(a).

As can be seen in the figure, even under a high level of congestion, the algorithm achieves a delivery rate of over 85\% in both of the test networks. Both networks exhibit a similar degree of degradation under the increasing load. Interestingly, in the Flickr dataset 10 pairs of nodes performed worst that 50 pairs. We believe that this is down to one of these 10 pairs of nodes not having a stable path between them. 

In both networks, the average delay for received messages is 1 day. This remains constant as the load increases. What does decrease, however, is the number of duplicate copies of a message that the destination receives. In Flickr for 100 pairs each destination receives on average 27 copies of each message. For 1000 pairs, this reduces to 10. In the BA model network, the number of duplicates reduces from 7.1 to 2.7.

\subsection{Effects of Black-holing}
A possible active attack against this system is for nodes in the system to be removed (either by the social network or insiders). The node does not have to be physically removed from the system, rather they can simply stop participating in the routing protocol. It will take some time for the delay distributions held by neighbours for that node to reflect this change in behaviour, meaning the pull decision, which may include the convolution of the removed node, could be made assuming a neighbour is active when it is not. This will have greatest affect when the node is a key link for many nodes. The goal of this attack is to perform a denial-of-service by breaking the paths between nodes. 

We test the impact of this attack by removing an increasing percentage of nodes from the network half way through the routing simulation. By removing the nodes from the graph, this has the same effect as the nodes no longer participating (due to the pull model). We use two strategies for removing nodes, choosing nodes at random, and choosing high degree nodes. We use high degree nodes as even though their is little correlation between upload behaviour and the degree of nodes, high degree nodes are still important for routing as they statistically will view more messages than others. High degree nodes are also likely to be the main connection to the rest of the network for some nodes, meaning their removal may prevent other nodes from communicating.

We use 100 pairs of nodes each communicating 1 message per month. We do not remove nodes that are in one of these pairs. The results of this experiment are shown in Figure~\ref{fig:performancey}(b).

As can be seen in the figure, in the face of the random removal of nodes, both networks are very resilient to even 50\% of the nodes being removed from the system, losing only 2-3\% of the delivery rate. The exception is for 40\% of the Flick network being removed, this has a slightly larger effect. We believe that this is due to the randomness -- the 40\% of nodes chosen in that test contained a larger percentage of nodes that were key to routing between the chosen pairs (the pairs were kept constant across all test but the nodes to remove were randomised for every test). 

Removing high degree nodes proved to be the most effective approach at causing a reduction in network performance, with both networks showing a drop down to 52\% and 56\% delivery rate when the 50\% highest degree nodes were removed. This is expected, as we have mentioned the high degree nodes, while not necessarily being high uploaders, are still important in the network for providing a connection point to many nodes. This effect is visible in the sparser Barabasi network; the delivery rate drops at a higher rate before levelling out due to more nodes being disconnected from the network completely. 

In both networks, and for both tests, the delay remains unchanged at \~{}1 day per message (that was delivered). The main noticeable effect, other than the reduction in delivery rate, is the reduction in the number of duplicate messages received by the destination, as much as 15\%. As we have already shown there is enough duplication, however, that we can tolerate the random node removal, and the messages that do arrive with the high degree nodes removed still arrive quickly.
\subsection{Path Consistency}
\begin{figure*}[!ht]
\centering
  \subfloat[Minimal entropy pair]
                { \includegraphics[width=0.3\textwidth]{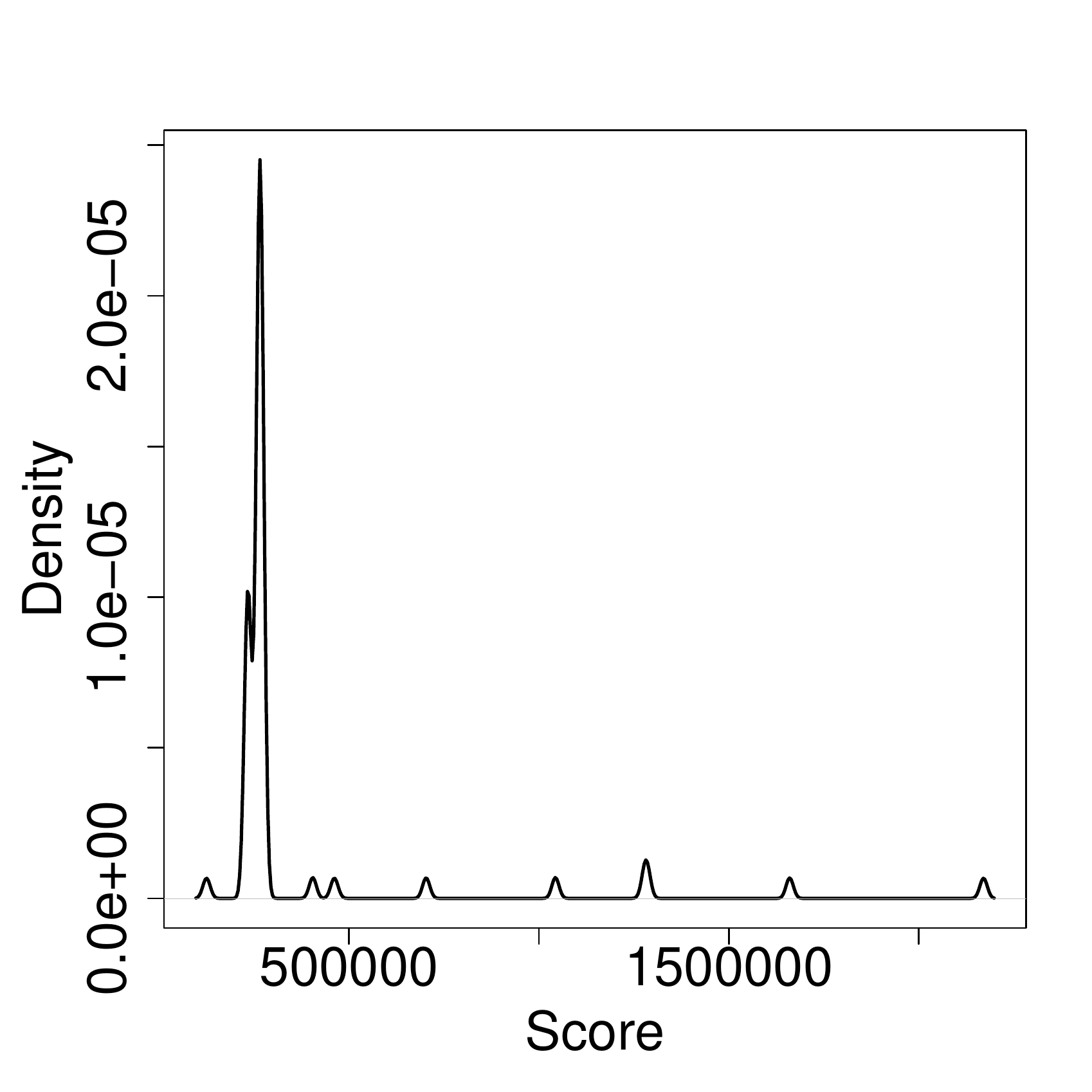}}
 \subfloat[Maximal entropy pair]
         { \includegraphics[width=0.3\textwidth]{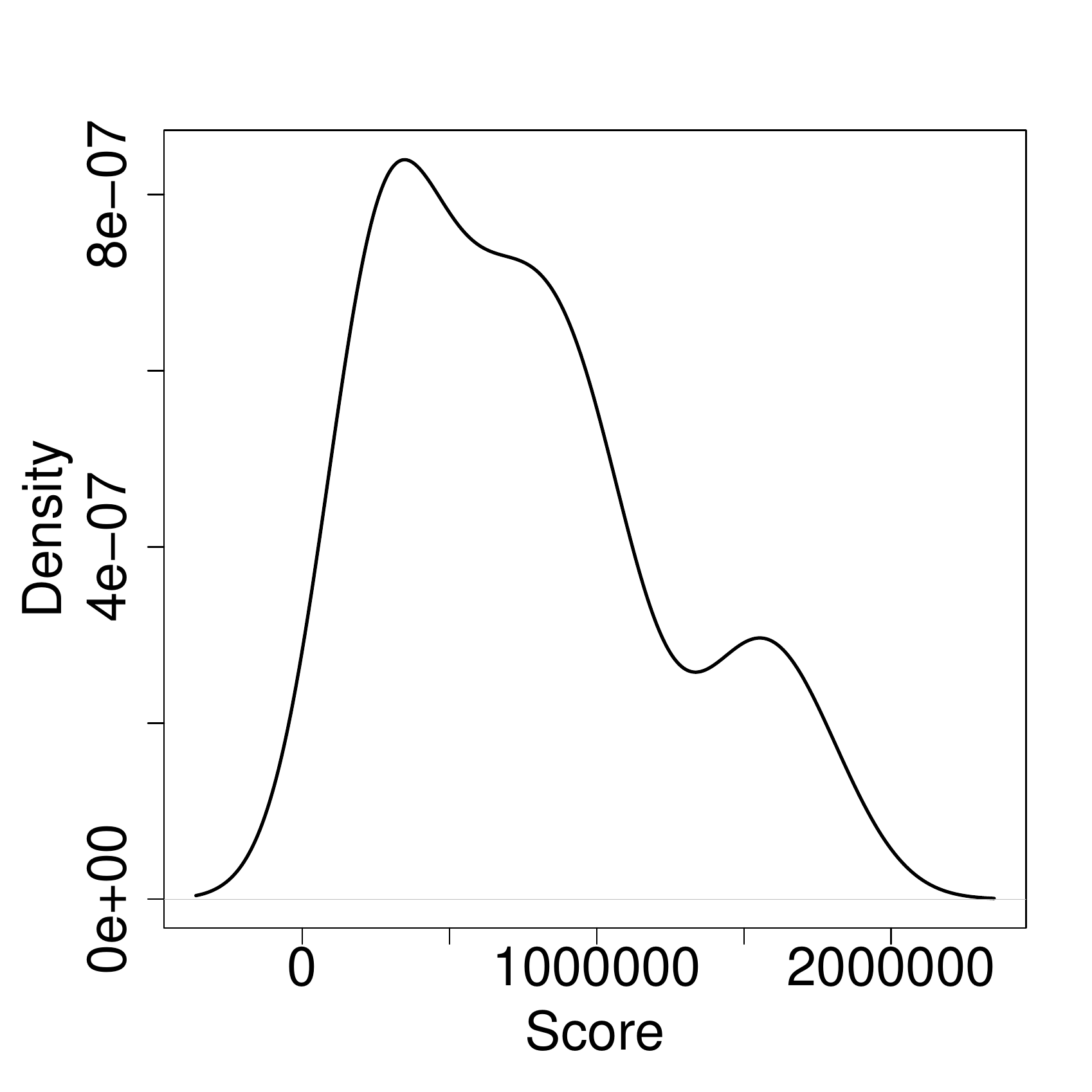}}
   \subfloat[Distribution of pair entropies]
         { \includegraphics[width=0.3\textwidth]{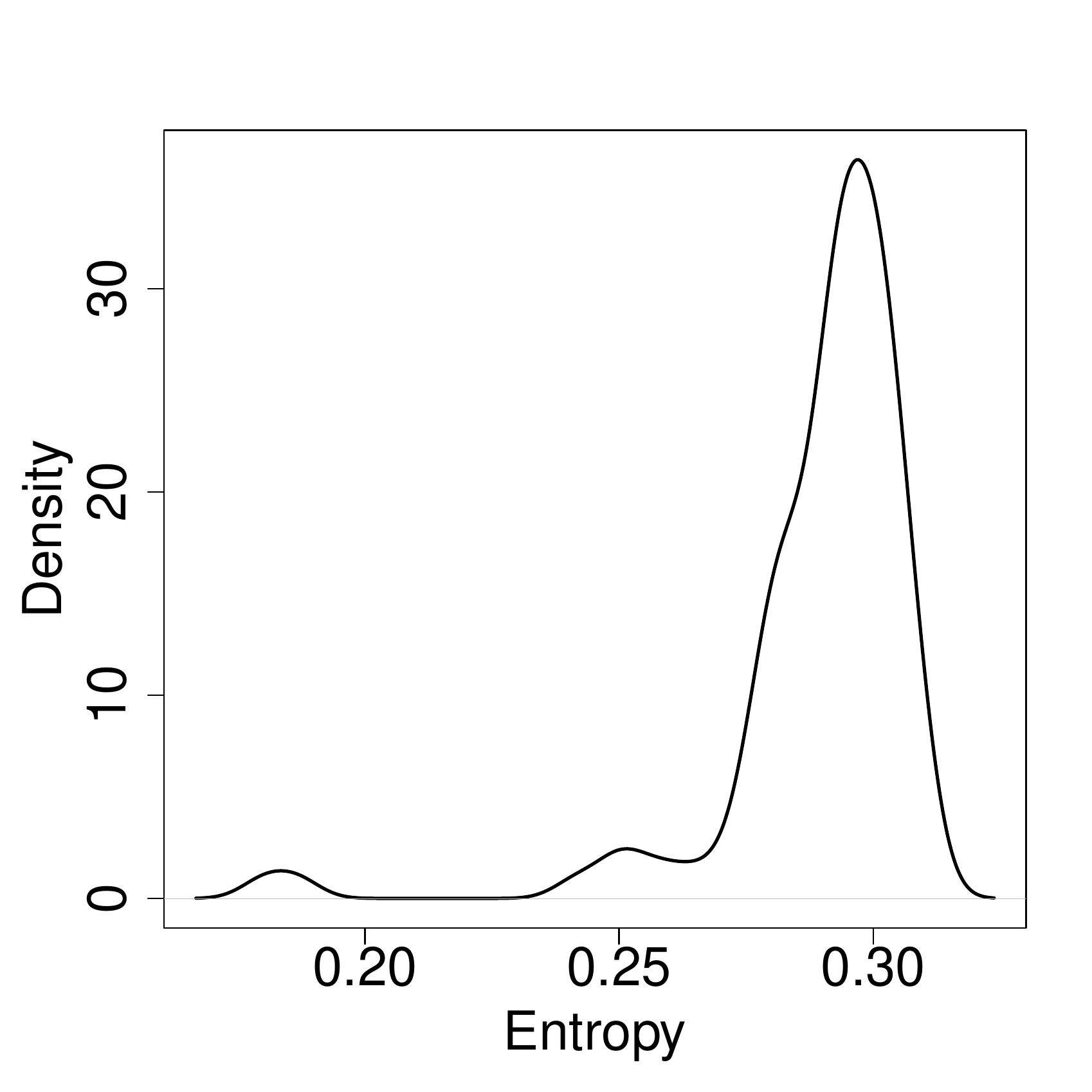}}
  
 \caption{Consistency of path quality}
\label{fig:entropy}
\end{figure*}

It is important that the paths taken by messages between pairs of nodes maintain their quality over time. We measure the quality of a path using the CDF score as defined in equation~\eqref{eq:cdfscore}, applied on the convolution of all nodes' distributions on the path. We compute the score for the first path that results in each message being delivered during the simulation (maximum 64 per pair of nodes). Due to the differing lengths of paths, we extend each path CDF such that all CDFs are of an equal length (by adding 1s). We then compute the Shannon entropy ($H(X) = -\sum_{i} {P(x_i) \log_b P(x_i)}$) of these scores for each pair of nodes, and produce a distribution of these values. Figure~\ref{fig:entropy} provides the score distributions of the pairs with the lowest and highest entropy, and the distribution of entropies across all pairs. 

Figure~\ref{fig:entropy}(c) shows that a majority of the pairs have an entropy of scores of approximately 0.3. This is a relatively low entropy value, meaning that, over time, the quality of the path does not exhibit much variation. This is indicative of the same paths being used repeatedly if they are available (all nodes uploading).

%% file: analysis.tex
\section{Analysis}\label{sec:analysis}

The primary goal of our system is to achieve indistinguishability. We
also set out to provide a system that is resilient against certain
insider attacks.

\subsection{Attacks on Communication Channel}
The system as described does not create any extra traffic, or introduce any new communication endpoints --- this is the basis for indistinguishability in our system. All communication is achieved using the user behaviour that already ``exists'' (the user would still upload images without the hidden channel). Traditional traffic classification methods would not be able to detect unusual behaviour. In addition, the use of good-listing or bad-listing would be ineffective. In countries where internet censorship does not occur, access to social networks is open. In places where social network access is not permitted, there is often a local, closed social network that is in use. The system can easily be transferred amongst social networks.

\subsubsection{Attacks on Routing}
We have shown that, under {\em targeted attacks} where insiders are able to remove the top 50\% highest degree nodes, the delivery rate can still maintain a level above 50\%, with no noticeable impact on latency. Under {\em random attacks} --  nodes are attacked at random --- the effect is a reduction in delivery rate of just a few percent. This is due to the large amount of disjoint paths taken by messages, unless every path is disrupted, delivery can still occur.

\subsubsection{Denial of Service Attacks}

Following our threat model, we assume that the OSN would be able to perform Denial-of-Service attacks on the system by destroying the steganographically hidden information. This could be done, for example, by applying heavy compression on an image (that even compression resistant steganography algorithms cannot survive) or by applying a transformation to the image in order to invalidate any hidden data. Facebook, for example, already applies heavy compression to all uploaded images. Any further compression will have a visible impact on image quality, as would any image transformations. 
Obviously, a country can completely block access to the social network itself. This is similar to the issue of a censor blocking access to mixes and relays in anonymous communication networks.

Unfortunately there is very little that can be done to avoid DOS
attacks. The redundancy provided by the routing algorithm helps to
prevent local DOS attacks, or in the situation where the adversary
only tampers with a subset of images at random. Past this, it is
important to make use of a steganography algorithm that provides
compression resistance which will prevent the attacker from simply
re-compressing images.

\subsubsection{Attacks on Steganography}
The major risk to the unobservability provided by the system is that
of steganalysis.

For a person that is already under targeted surveillance, steganalysis
could be applied to all images uploaded by that person and their
neighbours in an attempt to identify if they are taking part.

The fact that all image payloads are encrypted will hinder
steganalysis attempts. Assuming that steganalysis can be applied, the
attacker will also have to reliably identify cipher-texts within the
image and differentiate them from random data. This will prove
difficult if they do not know the cipher-texts that are being
sought. Further to this, using keyed steganography algorithms with
individual keys will also hinder steganalysis attempts.

As mentioned in section~\ref{sec:stego}, there are steganography
algorithms that are considered statistically undetectable when used
with small payloads. A common mistake when using steganography is the
use of large payloads which (a) makes traffic analysis on the images
possible and (b) renders the channel vulnerable to compression-based
attacks. We have avoided both of these issues in our design.

If a user of the social network is found to be using the system,
unlinkability and message confidentiality will be maintained, meaning
the user will be reduced to the security level provided by previous
anonymity systems such as Tor.

\subsection{Resistance to Insider Attacks}
Our design ensures sender-receiver unlinkability~\cite{PH00} in
multiple cases. First, a single compromised node does little damage:
there is no sender address included with message blocks, so an
attacker node has no knowledge of the source. As the destination
address is encrypted (and subsequently universally re-encrypted) with
the neighbourhood key of the destination only nodes with the private
neighbourhood key of the destination know where it is going. Second,
the use of re-encryption at each node along the route offers traffic
analysis resistance against passive colluding attackers by ensuring
bit-wise unlinkability between incoming and outgoing message
blocks. Finally, resistance against active collusion including replay
attacks and tagging attacks is ensured (during re-encryption) by
raising input ciphertexts to a random number from $Zq$. Due to this
operation, replayed message blocks will not result in identical output
message blocks. Similarly, any changes (tagging) to an input block
will be blinded during re-encryption, hence tagged inputs will not be
recognisable at the output. The inherent protection provided by making
use of a social network against Sybil attacks also increases the
effort required by an attacker to inject malicious nodes in the first
place.

\subsection{Design Limitations}
\subsubsection{Ciphertext lengths}
There is a major downside to using universal re-encryption, in that
for every $k$ bits of data, $4k$ bits of ciphertext are required to be
transmitted. However, this is justified given the high risk of insider
attacks. Despite these overheads, Blindspot remains a usable system
for low-bandwidth anonymous communications where indistinguishability
is required to avoid censorship.

\subsubsection{Insider key leakage}
The requirement of neighbours being able to identify messages destined to neighbours opens the system to a form of insider attack. If a neighbour of a node is compromised, they can leak the decryption key which will enable any node also under the attacker control to identify messages destined for that destination. The message contents, however, are not at risk (as only the recipient has the decryption key). 

Making use of key-based steganography goes some way in alleviating this problem. If the steganography is open (meaning no key or a global key is used) the global adversary would be able to capture all images, extract all of the embedded data and decrypt destination addresses. If individual keys are used for steganography which are only shared with neighbours, only other insiders will be able to identify messages as the messages will only be hidden using that key at the node for which they are uploaded. At other points in the path a different steganographic key would be used meaning that the embedded messages are not retrievable.

\subsubsection{Out-of-band key distribution}
Currently, the system requires a large amount of out-of-band key distribution. As the system is fully decentralised, low bandwidth and high latency standard key distribution protocols are not optimal. This problem could in part be solved through the use of identity-based cryptography, although this introduces the need for a key escrow which introduces a communication endpoint.

%% file: background.tex
\section{Background and Related Work}\label{sec:background}
\subsection{Unobservable Communications}

There have been many attempts at providing unobservable communication on the
internet.  One common approach is to use steganography in tandem with web-based
storage systems to provide methods of communication. Unfortunately, none of
these techniques provide anonymity and indistinguishability and hence are of
little use in our threat context.

CovertFS~\cite{Baliga:07:covertfs} is a prototype system for
providing a covert file store that stores data (and file structure) in images
stored on photo sharing sites. The data can be shared amongst users of the
system, with a goal of achieving plausible deniability - i.e. a user can deny
that they are using the system (which is reliant on the resistance of the
steganography algorithm used to steganalysis). 

Collage~\cite{Burnett:2010:CAC} is another system that follows a similar approach. In Collage, data is hidden within media which is uploaded to a user-generated content host. The receiving party then accesses the media by performing some task (such as accessing a URL or searching for specific terms), downloads it and extracts the hidden data. The key point is that the content host is interchangeable -- i.e. if one service is blocked another can be used with minimal effort. This is all in full view of the censor, who sees only the transfer of the media file from the uploader to the media server (a legal operation in most states), and the download/access of the file by the receiver (again, a legal, innocent operation in most cases). The censor is assumed to have the ability to collect full traffic information on a target, and disrupt communications. An extension to the system is proposed by Invernizzi et al.~\cite{Invernizzi:2013:MBS}, in which data is hidden within images on blog posts. The system is different however, as it allows for any blog to be used on any provider, meaning that there are no single entry points (e.g. Flickr or Facebook) to block. The rendezvous point is found using a search engine. The deniability arises from the fact that every minute hundreds of thousands of images are uploaded in blog posts, and so it is infeasible for the censor to either run steganalysis on every image, or block every blog provider. 

Collage does not work under our threat model. Under targeted surveillance, the attacker can easily isolate Collage communications and break the user's anonymity. This down to the fact that collage introduces new endpoints into communication (the source and receiver(s) connecting to the intermediary), which will appear in traffic logs (including GCHQ collected Tempora data)  and may differ from normal behaviour.

The above systems are high-latency systems designed for one-off transmissions of data. There are also systems that use steganography for lower-latency communications. For example, Huang et al.~\cite{huang:08:stegviop} propose a system for transmitting data by hiding information within VoIP packets. Craig Rowland~\cite{rowland:97:ip} presents a number of methods for embedding data within TCP/IP packet headers by modifying particular fields (those that are optional, unused and are unlikely to be changed in transit). Ahsan and Kundar~\cite{ahsan:02:tcpip} also hide data within the headers of TCP/IP packets, but also propose a second method of data hiding by changing the ordering of packets. This can achieve a bandwidth $log(n!)$ bits (where $n$ is the number of packets, for example 25 packets provides around 87 bits of capacity). This method could be discovered, however, if all packets consistently arrive (or are sent) out of order using simple traffic analysis. Modified TCP/IP headers are also trivial to detect if the header fields contain non standard (default) values.

An alternative to steganography is protocol mimicking, in which attempts are made to make one traffic protocol appear as another. Due to its high detectability rate but high usage rates, Tor is often the subject of work to try and make it less detectable. For example, StegoTorus~\cite{weinberg:2012:stegotorus} and SkypeMorph~\cite{moghaddam:2012:skypemorph} both attempt to hide Tor traffic by making it appear as VoIP (in particular Skype) traffic. The systems attempt to produce protocols that can carry Tor traffic, but exhibit the characteristics of the Skype protocol, including timing and packet headers. CensorSpoofer~\cite{wang:2012:censorspoofer}  is designed for obfuscated web browsing by decoupling the upstream and downstream channels of a HTTP session. HTTP requests are sent to the server over a low capacity channel such as email or instant messaging, and the server responds to the client  mimicking UDP-based VoIP traffic, by emulating the traffic from a dummy P2P host, more specifically SIP-based VoIP. All three of these systems have been proven broken~\cite{houmansadr:2013:parrot}, specifically because they are unable to fully emulate the protocols they mimic, meaning that their usage can be detected by looking for unusual behaviour (for example incorrect error handling or malformed headers).

Houmansadr et al.~\cite{freewave} propose Freewave, a system for converting IP traffic into VoIP traffic for circumventing internet censorship. IP packets are converted into acoustic signals, which are then transmitted to a Freewave server, which converts the VoIP stream back into IP packets and releases them into the uncensored internet, acting as a proxy. The system uses encryption to make sure that the Freewave VoIP packets are indistinguishable from from background packets. Due to the use of Skype supernodes in routing, the source is unlinkable from the server, so even if the censor knows the servers identity, they will not automatically know the identities of end users. The VoIP operator, however, may know who communicates with the Freewave server, so it is assumed that a provider who does not collude with the censor is used (a weak assumption). The authors state that the system is resilient to traffic analysis as the calls follow the same traffic patterns as a normal VoIP call, but there is one small issue in this assumption. If Freewave is used for web-browsing rather than normal speaking, then the frequency and timing of calls may be an indicator of the systems use (if it different from normal usage patterns of users). The call frequency and use pattern for a user who is actually using the service for web browsing rather than calling could be substantially different to that of the user before they started browsing, or before the started using the service for browsing.

\subsection{Decentralised Routing over Social Networks}

Fully distributed anonymous routing precludes the use of
design options such as onion routing or its semi-centralised
peer-to-peer variants. This is due to their dependency on the
availability of a central directory service. Even with a fully
distributed directory service, we cannot make conventional onion
routing work due to the capacity constraints of each node; deploying
such a directory service would almost certainly exhaust the meagre
network bandwidth available leaving little opportunity for efficient
anonymous communication.

Significant prior art exists on decentralised routing protocols in the
context of peer-to-peer adhoc networks. 
 
Gonzalez et al.~\cite{gonzalez:nature08} proposed
routing algorithms that leveraged human mobility patterns. Their algorithm was
design to handle edges which were available according to an exponential
distribution. These findings support those made earlier by Clauset and
Eagle~\cite{clauset:dimacs07} on a smaller dataset of $66$
individuals. Chaintreau~\cite{chaintreau:conext07} observed that in graph
topologies based on human mobility paths of $O(log\ n)$ hops exist. These
findings are strongly reminiscent of the characteristics of `small-world'
networks which are thought to closely resemble human social
networks~\cite{WS98}.

We can also gain some understanding of these networks by several
attempts to construct routing protocols. One of the early
proposals~\cite{epidemic} was {\em epidemic routing} which is
essentially flooding. Other studies~\cite{williamson:iwsos09} tested the
effects of flooding on popular datasets such as the MIT Reality mining
dataset\footnote{http://reality.media.mit.edu/dataset.php}. This
dataset consists of the mobility patterns of 100 individuals over a
nine-month period. The authors found (see Fig.~2
of~\cite{williamson:iwsos09}) that (owing to the periodicity and
frequency of human mobility) the message delivery rates were
time-invariant; when the time-to-live is between three to five days,
50-60\% of the messages were delivered on the average when all pairs
of nodes simultaneously communicate.

While epidemic routing(flooding) makes few assumptions about graph
structure, it causes congestion for well understood reasons which
ultimately results in lower message delivery rates since the network
is flooded with duplicates. For these reasons, a number of schemes
that use resources more efficiently have been proposed. Spyropoulos et al.~\cite{spray} produced a routing algorithm in which nodes forward limited copies of the data to nodes that are encountered within the network. If a node is directly connected to the destination, it sends to the packet. This method relies heavily on the mobility of nodes to enable the data packets to reach their destination. This approach requires the use of a decreasing copy count. The copy count would leak a large amount of information about the source of the message (a node on the path knows how far the source is from them in hops). The lack of mobility also vastly reduces the effectiveness of the algorithm in a static network. Daly and Haar~\cite{daly:mobihoc07} proposed using a centralised path discovery technique to find viable shortest paths. This involves using an $O(n^2
log n)$ algorithm to find and route through the most
betweenness-central nodes in a graph, using Freeman's~\cite{F77}
definition of betweenness centrality. The authors showed that their
scheme performs almost as well as the basic flooding algorithm. Hui et
al.~\cite{bubblerap} found that betweenness centrality and degree
centrality are highly correlated and therefore using a local
centrality metric to drive route selection can also result in
performance (delivery ratios) close to that achieved by flooding. Both of these systems are unsuitable in this system, however, as they both require knowledge of both the destination and the betweenness of nodes with relation to it. This information would break unlinkability.

Mittal et al.~\cite{xvine} propose X-Vine, a system for providing defence against Sybil attacks on distributed-hash table based systems. The proposed design applies a social network overlay to the distributed hash table (based upon Chord). The Chord routing algorithm is used as normal at a high, but rather than route directly between Chord points, data is routed through a subsection of the social network. The paper has a goal of low control data overhead, but it is still far above an acceptable level for our setting. While this level is suitable for networks without the stringent bandwidth limitations of ours, it would use up too much valuable capacity in the low-bandwidth environment. 

Blindspot has some similarities to Crowds~\cite{Reiter:1998:crowds}. Within Crowds, traffic is randomly routed between members of a crowd until one member decides to forward the message to the server. Each member on a path does not know if the preceding member was the originator of the traffic. The routing provided by Blindspot also provides this property; a node is unable to tell if a message it receives or forwards originated from the node it pulled the message from. The major difference, however, is that in Crowds the goal is to communicate with a server that any node can establish a direct connection to (enabling the traffic to reach the destination), whereas in Blindspot a path needs to be found between the source and destination on the underlying social graph.

%% file: conclusion.tex
\section{Conclusion}\label{sec:conclusion}
\subsection{Future Work}
One of the major pieces of future work will be to produce a prototype of Blindspot and to perform a limited beta (for example limited to the author's department). The software will be of the form of a browser plugin that will intercept user's uploads and perform the system operation automatically with minimal user input. There is also scope for creating a version of the software for smartphones due to their increasing use as cameras that upload images directly to social networks. The main goal of the prototype will be to perform a usability study of the system in operation. 

One area that needs to be explored in the future is the issue of whether or not users who are participating in the Blindspot (or similar) system may change their uploading behaviour in order to support the system. While this is good for the system's operation, the change in behaviour may be an indication of the systems use. Exploring whether or not users will change their behaviour noticeably while using systems such as Blindspot will be important for any system that aims to achieve unobservability.  
Finally, currently the system design focuses on the system being present on a single social network. In reality, users are often members of many social networks. The users may behave differently on different networks (a person could well upload more images on their Flickr account that on their Facebook account), but have overlapping social graphs. To this end, the system design could be adapted in such a way that the users entire social graph is used (built from all of their social networks). This could also allow the system to make use of cover texts other than images (including text and video) in order to increase the network capacity. 
\subsection{Conclusion}
Evident use of anonymous communication networks can present real-world risks to
the lives of users in some countries. It can also allow a censor to wholly block
anonymous communications. In this paper, we have proposed Blindspot which is
specifically designed to address the needs of protest organisers and others. Its
goals are to provide indistinguishability and unobservability against the
network infrastructure provider. We have shown that leveraging social behaviour
of users is an attractive approach to achieve these goals. In conclusion, our
main contribution is a novel decentralised routing protocol that provides high
levels of performance, given the constraints, while providing anonymity against a global passive adversary. It  leverages trust relationships on a social network to contain node misbehaviour and offers some protection against insider attacks such as blackholing.